\documentstyle[11pt]{article}
\input epsf
\def\sdtimes{\mathbin{\hbox{\hskip2pt\vrule
height 4.1pt depth -.3pt width .25pt\hskip-2pt$\times$}}}

\begin{document}
\thispagestyle{empty}
\baselineskip 20pt
\rightline{KIAS-P00045}
\rightline{NBI-HE-00-32}
\rightline{SOGANG-HEP 274/00}
\rightline{{\tt hep-th}/0007140}

\def\sdtimes{\mathbin{\hbox{\hskip2pt\vrule
height 4.1pt depth -.3pt width .25pt\hskip-2pt$\times$}}}

\def\tr{{\rm tr}\,}
\newcommand{\IR}{\relax{\rm I\kern-.08em R}}
\newcommand{\IZ}{\relax\ifmmode\mathchoice
{\hbox{\cmss Z\kern-.4em Z}}{\hbox{\cmss Z\kern-.4em Z}}
{\lower.9pt\hbox{\cmsss Z\kern-.4em Z}}
{\lower1.2pt\hbox{\cmsss Z\kern-.4em Z}}\else{\cmss Z\kern-.4em
Z}\fi}
\newcommand{\beq}{\begin{equation}}
\newcommand{\eeq}{\end{equation}}
\newcommand{\beqn}{\begin{eqnarray}}
\newcommand{\eeqn}{\end{eqnarray}}
\newcommand{\n}{\nonumber}
\newcommand{\ct}{\cite}
\newcommand{\bde}{{\bf e}}
\newcommand{\balpha}{{\mbox{\boldmath $\alpha$}}}
\newcommand{\bsalpha}{{\mbox{\small\boldmath$alpha$}}}
\newcommand{\bbeta}{{\mbox{\boldmath $\beta$}}}
\newcommand{\btau}{{\mbox{\boldmath $\tau$}}}
\newcommand{\blambda}{{\mbox{\boldmath $\lambda$}}}
\newcommand{\bepsilon}{{\mbox{\boldmath $\epsilon$}}}
\newcommand{\bphi}{{\mbox{\boldmath $\phi$}}}
\newcommand{\bpi}{{\mbox{\boldmath $\pi$}}}
\newcommand{\bX}{{\mbox{\boldmath $X$}}}
\newcommand{\ggg}{{\boldmath \gamma}}
\newcommand{\ddd}{{\boldmath \delta}}
\newcommand{\mmm}{{\boldmath \mu}}
\newcommand{\nnn}{{\boldmath \nu}}

\newcommand{\bra}[1]{\langle {#1}|}
\newcommand{\ket}[1]{|{#1}\rangle}
\newcommand{\sn}{{\rm sn}}
\newcommand{\cn}{{\rm cn}}
\newcommand{\dn}{{\rm dn}}
\newcommand{\diag}{{\rm diag}}

\vskip 2cm

\centerline{\LARGE\bf The CP(n) Model on Noncommutative Plane}

\vskip 1.5cm
\centerline{\large\it  Bum-Hoon Lee$^{\;a\;}$\footnote{bhl@ccs.sogang.ac.kr},
 Kimyeong Lee$^{\;b\;c\;}$\footnote{klee@kias.re.kr},
and Hyun Seok Yang$^{\;b\;}$\footnote{hsyang@physics4.sogang.ac.kr} }

\vskip 1mm \centerline{$^{a\;}$Department of Physics, Sogang
University} \centerline{Seoul 121-742, Korea}
\vskip 1mm

\centerline{$^{b\;}$School of Physics, Korea Institute for Advanced Study}
\centerline{207-43 Cheongriangri-Dong, Dongdaemun-Gu}
\centerline{Seoul 130-012, Korea}
\vskip 1mm

\centerline{$^{c\;}$The Niels Bohr Institute}
\centerline{Blegdamsvej 17, DK-2100 Copenhagen, Denmark}

\vskip 3mm

\vskip 1.cm
\begin{quote}
{\baselineskip 16pt We construct the consistent $CP(n)$ model on
noncommutative plane. The Bogomolny bound on the energy is
saturated by (anti-)self-dual solitons with integer topological
charge, which is independent of their scaling and orientation.
This integer quantization is satisfied for our general solutions,
which turns out regular everywhere. We discuss the possible
implication of our result to the instanton physics in Yang-Mills
theories on noncommutative ${\bf R}^4$.}
\end{quote}

\newpage
\setcounter{footnote}{0}

\section{Introduction}

Quantum field theory on a noncommutative space has been proved to
be useful in understanding various physical phenomena, like as
various limits of M-theory compactification \ct{cds,ho}, low
energy effective field theory of D-branes with constant
Neveu-Schwarz $B$-field background \ct{aas,sw}, and quantum Hall
effect \ct{bell}. Although noncommutative field theories are
non-local, they appear to be highly constrained deformation of
local field theory. Thus it may help understanding non-locality at
short distances in quantum gravity.

Noncommutative field theory means that fields are thought of as
functions over noncommutative spaces. At the algebraic level, the
fields become operators acting on a Hilbert space as a
representation space of the noncommutative space. Since the
noncommutative space resembles a quantum phase space (with
noncommutativity $\theta$ playing the role of $\hbar$), it
exhibits an interesting spacetime uncertainty relation, which
causes a $UV/IR$ mixing \ct{mrs} and a teleological behavior
\ct{sst}.  Also, for nonzero $\theta$, there can be
nonperturbative effects in the form of soliton solutions even at
the classical level and it could not have a smooth limit. Indeed,
several such solutions have recently appeared
\ct{naka,ns,ky,tera,bn,furu,kly,hh,gms,gn,poly}.

In this Letter we will show the existence of nonperturbative
solutions in the $CP(n)$ model on noncommutative two plane. The
$CP(n)$ model, even though it consists only of scalar fields,
enjoys local gauge invariance and exhibits many similarities to
instantons in four-dimensional Yang-Mills theory, such as the
existence of self-dual soliton solution  with scale and
orientation parameters \ct{egp,ddl}. In addition, it has many
applications to condensed matter systems \ct{bp}.  However, since
the $CP(n)$ model is relatively simpler than four-dimensional
noncommutative Yang-Mills theory, it will be very useful ``toy
model'' to investigate various questions of noncommutative gauge
theory if the ordinary $CP(n)$ model can be generalized to
noncommutative space. We will demonstrate here it is the case.

In Section 2, we construct the consistent $CP(n)$ model on
noncommutative plane. In Section 3, the Bogomolny bound on the
energy is considered and it is shown that it is saturated by
(anti-)self-dual solutions.  In Section 4, we solve the soliton
solutions for the (anti-)self-dual equations and show that their
topological charges, which are independent of their scaling and
orientation, are quantized. However, we point out that this
integer quantization is satisfied only for the field
configurations without any singularity in the commutative sense.
Also we argue that our solution is the most general BPS solution.
In Section 5, we summarize our results with a brief discussion of
the possible implication to the instanton physics in Yang-Mills
theories on noncommutative ${\bf R}^4$ \ct{ns}, including the
problem for noncommutative instanton solutions discussed in
\ct{kly}.

\section{CP(n) Model}

We consider the (2+1)-dimensional field theory whose space is
noncommutative two plane. The coordinates $x,y$ of this noncommutative
plane satisfy the relation
\begin{equation}
[x,y] = i \theta
\label{comm}
\end{equation}
with $\theta>0$. This noncommutative plane has not only the
translation symmetry but also rotational symmetry. One can see that
the parity operation $(x,y)\rightarrow (x,-y)$ is broken on
noncommutative plane. The classical field on this noncommutative  space is an
element $\Phi(t,x,y)$ in the algebra ${\cal A}_\theta$ defined by
$x,\, y$ with the relation $(\ref{comm})$.

Introduce the complex coordinates
\beq
z = \frac{x+iy}{\sqrt{2}}, \;\; \bar{z} = \frac{x-iy}{\sqrt{2}},
\eeq
which satisfy
\beq
[z,\bar{z}] = \theta>0.
\eeq
This commutation relation is that of the creation and annihilation
operators for a simple harmonic oscillator and so one may use the
simple harmonic oscillator Hilbert space ${\cal H}$ as a
representation of (1).  The ground state is $|0>$ such that $z|0>=0$
and $|n> = \bar{z}^n/\sqrt{\theta^n n! }|0>$ so that
\beq
z|n> = \sqrt{\theta n} |n-1> , \;\;\; \bar{z}|n> = \sqrt{\theta(n+1)}|n+1>.
\eeq
The integration over noncommutative two plane becomes the trace
over its Hilbert space, which respect the translation symmetry:
\beq
\int d^2x {\cal O}  \rightarrow  {\rm Tr} {\cal O} =
2\pi \theta \sum_{n\ge 0}
<n|{\cal O} |n>.
\eeq

The $CP(n)$ manifold  is defined by an $(n+1)$-dimensional complex vector
$\Phi=(\phi_1,\phi_2,...,\phi_{n+1})$ of unit length with the
equivalence relation under the overall phase rotation $\Phi \sim
e^{i\alpha} \Phi$ \ct{egp,ddl}.
This complex projective space of real  dimension
$2n$ is equivalent of the coset space $U(n+1)/U(1)\times U(n)$. (It
is quite straightforward to generalize our consideration here to the
general Grassmanian models with the manifold $G(n,m) =U(n)/U(m)\times
U(n-m)$ \ct{mac}.)

Here we are interested in the $CP(n)$ model on the noncommutative plane.
Thus the field variable $\Phi(t,x,y)$ becomes an operator acting
on ${\cal H}$. The spatial derivatives are
\begin{equation}
\partial_x\Phi= i\theta^{-1}[y,\Phi],\;\;\; \partial_y\Phi= -i\theta^{-1}
[x,\Phi].
\end{equation}
The natural Lagrangian for the $CP(n)$ model turns out to be
\beq
L_1 = {\rm Tr} \left( \partial_\mu  \Phi^\dagger \partial^\mu \Phi
+ (\Phi^\dagger \partial_\mu \Phi)( \Phi^\dagger \partial^\mu \Phi) \right)
\label{lag1}
\eeq
with the constraint
\beq
\Phi^\dagger \Phi - 1 = 0.
\label{norm}
\eeq
This theory has a global $U(n+1)$ symmetry and a local $U(1)$ symmetry
\beq
\Phi(x) \rightarrow \Phi(x) g(x)
\label{gauge1}
\eeq
which removes the degrees of freedom for the overall phase of
$\Phi$. The $U(1)$ gauge transformation acts on the right side,
which leaves the constraint (\ref{norm}) invariant.  This ordering
of the gauge transformation is the key point which makes the whole
model work.

This Lagrangian with the constraint (\ref{norm}) can be rewritten as
\beq
L_2= {\rm Tr} \left( D_\mu \Phi^\dagger D^\mu \Phi
 + \lambda (\Phi^\dagger \Phi - 1)
\right)
\label{lag2}
\eeq
with
\beq
D_\mu \Phi = \partial_\mu \Phi -i \Phi A_\mu,
\eeq
where $A_\mu(x)$ is the $U(1)$ gauge field without its kinetic
term and $\lambda(x)$ is a Lagrangian multiplier to incorporate
the constraint (\ref{norm}). This Lagrangian is invariant under
the local gauge transformation defined by (\ref{gauge1}) and
\beq
A_\mu(x) \rightarrow g^\dagger  A_\mu g -ig^\dagger \partial_\mu g.
\label{gauge2}
\eeq
As there is no gauge kinetic term, one can solve the $A_\mu$ equation to
get
\beq
A_\mu = -i \Phi^\dagger \partial_\mu \Phi
\label{aeq}
\eeq
which shows that the gauge transformations (\ref{gauge1}) and
(\ref{gauge2}) are consistent with one another. After using
Eqs.~(\ref{norm}) and (\ref{aeq}), the second Lagrangian (\ref{lag2})
becomes the first Lagrangian (\ref{lag1}), as it should be.

Note that $\Phi^\dagger D_\mu \Phi =0$ and the field strength is
\beqn
F_{\mu\nu} &=& \partial_\mu A_\nu - \partial_\nu A_\mu + i[A_\mu,A_\nu] \\
 &=& -i(D_\mu \Phi^\dagger D_\nu \Phi - D_\nu \Phi^\dagger D_\mu \Phi)
\eeqn
which is the curvature tensor $[D_\mu,D_\nu]\Phi = -i \Phi F_{\mu\nu}$.
From the field equation for $\Phi$,
\beq
D_\mu D^\mu \Phi - \Phi \lambda = 0,
\eeq
we can deduce
\beq
\lambda = \Phi^\dagger D_\mu D^\mu \Phi = -D_\mu \Phi^\dagger D^\mu \Phi,
\eeq
and the field equation becomes
\beq
D_\mu D^\mu \Phi + \Phi (D_\mu \Phi^\dagger D^\mu \Phi)=0.
\eeq

\section{Energy  Bound }

As in the commutative case, the $CP(n)$ model on the
noncommutative plane has the Bogomolny energy bound. The conserved
energy functional becomes
\beqn
E &=& {\rm Tr} \left( D_0\Phi^\dagger D_0\Phi + D_i\Phi^\dagger D_i \Phi)
\right) \n \\
&=& {\rm Tr} \left( |D_0\Phi|^2 + |D_z\Phi|^2+ |D_{\bar{z}}\Phi|^2\right).
\eeqn
Similar to the commutative case, let us consider an inequality
\beq
{\rm Tr} \left\{ ( D_i \Phi \pm i \epsilon_{ij} D_j \Phi)^\dagger
(D_i \Phi  \pm i \epsilon_{ij}D_j \Phi)\right\} \ge 0.
\eeq
Expending this we obtain
\beq
{\rm Tr} (D_i\Phi^\dagger D_i \Phi ) \ge \mp i \epsilon_{ij} {\rm Tr}(
D_i \Phi^\dagger D_j \Phi ).
\eeq
The BPS bound on the energy is then
\beq
E \ge {\rm Tr } (D_0\Phi^\dagger D_0 \Phi)  + 2\pi |Q|,
\eeq
where the  $U(1)$ gauge invariant `topological charge' is
\beq
Q = -\frac{i}{2\pi} \epsilon_{ij}{\rm Tr} D_i\Phi^\dagger D_j \Phi =
\frac{{\rm Tr} F_{12}}{2\pi} .
\eeq
Contrast to the commutative case, there exists no conserved
topological current. Instead, the  current
\beq
J^\mu = \frac{1}{4\pi} \epsilon^{\mu\nu\rho} F_{\nu\rho}
\eeq
is covariantly conserved, $D_\mu J^\mu = 0$. However, this implies
that for the localized configurations, the topological charge $Q
={\rm Tr} J^0$ is conserved. In the complex coordinate,
\beq
Q = \frac{1}{2\pi} {\rm Tr} \left(|D_{z} \Phi|^2 - |D_{\bar{z}} \Phi|^2
\right).
\eeq

The energy bound is saturated by the configuration which is static in
time and satisfies the (anti-)self-dual equation \ct{ddl},
$D_i \Phi \pm i \epsilon_{ij}D_j \Phi = 0 $, or in the complex notation,
\beqn
&& D_{\bar{z}}\Phi=0 \;\;\; ({\rm for \;\;
self{\mbox{-}}dual\;\; case}\;\; Q>0), \\
&& D_z \Phi=0 \;\;\; ({\rm for\;\;
anti{\mbox{-}}self{\mbox{-}}dual\;\; case}\;\; Q<0).
\eeqn

\section{(Anti-)Self-dual Solitons}

To find the (anti-)self-dual configurations, let us try to parameterize
the field as follows
\beq
\label{cpn-coord}
\Phi = W /\sqrt{W^\dagger W} ,
\eeq
where $W$ is an $(n+1)$-dimensional vector. Since $\Phi^\dagger
\Phi=1$, locally one can choose finite $W$ such that
$\sqrt{W^\dagger W}(x)$ is invertible. We also introduce an
$(n+1)$-dimensional projection operator
\beq
P = 1 - W \frac{1}{W^\dagger W} W^\dagger ,
\eeq
whose kernel is one-dimensional space generated by $W$ vector. In
terms of this field variable, the Lagrangian (\ref{lag1}) becomes
\beq
L = {\rm Tr} \left( \frac{1}{\sqrt{W^\dagger W } } \partial_\mu W^\dagger P
\partial^\mu W \frac{1}{\sqrt{W^\dagger W}} \right),
\eeq
and the topological number is
\beq
\label{Q(W)}
Q = \frac{1}{2\pi}{\rm Tr}  \left\{ \frac{1}{\sqrt{W^\dagger W}}
  \left( \partial_{\bar{z}} W^\dagger P\partial_z W -
\partial_z W^\dagger P \partial_{\bar{z}} W \right)
\frac{1}{\sqrt{W^\dagger W}} \right\}.
\eeq
The operator in the trace can be regarded as the topological
density operator on the noncommutative space.  In terms of $W$
variable, the above Lagrangian in the commutative case has a local
scaling symmetry, $W \rightarrow W \Delta(x) $ with an arbitrary
scalar $\Delta(x)$. What is remarkable about the noncommutative
space case is that this scaling symmetry on the Lagrangian still
holds. The projection operator $P$ is independent of $\Delta$ and
so the Lagrangian and the topological number are still invariant.
However, the multicomponent field $\Phi$ is not invariant under
the local scaling on noncommutative case, contrast to the
commutative case. As far as the classical field theory is
concerned, we could choose the primary field to be $W$ instead of
$\Phi$, and regard the classical theory is invariant under the
local scaling. This would be crucial in finding the most general
solution.

The self-dual equation then becomes
\beq
D_{\bar{z}} \Phi = P( \partial_{\bar{z}} W )(W^\dagger W)^{-1/2} = 0 ,
\label{Weq}
\eeq
which is equivalent to $\partial_{\bar{z}} W = WV$ for arbitrary
scalar $V$ both for commutative and noncommutative cases. For
either cases the most general solution is
\begin{equation}\label{gsol}
W = W_0(z) \Delta(z,{\bar z}) \end{equation} with
$(n+1)$-dimensional holomorphic vector $W_0(z)$ and arbitrary
scalar function $\Delta(z,{\bar z})$. As we just argued in the
previous paragraph, this arbitrariness is a local scaling and can
be scaled away.

Let us consider the self-dual solutions in commutative case, which
is well studied before. We choose the scaling so that the
$(n+1)$th component of $W$ is chosen to be unity. Then, we get the
standard self-dual equation for the $n$-dimensional vector $w$
such that $W= (w,1)$
\beqn
&& \partial_{\bar{z}} w =0 \;\;\;\; ({\rm for \; self{\mbox{-}}dual}),\\
&& \partial_z w=0 \;\;\;\;
({\rm for\; anti{\mbox{-}}self{\mbox{-}}dual}).
\eeqn
The most general solution of the above self-dual equation should
be a $n$-dimensional vector whose components are holomorphic
functions. These solutions are characterized by its topological
charge $k$: the self-dual solutions carry positive integer charges
and the anti-self-dual solutions do negative integers. The general
self-dual solutions in commutative case are given in the
meromorphic form,
\beq w= \frac{1}{P_{n+1}(z)}(P_1(z), P_2(z), \cdots,P_n(z)),
\label{type1} \eeq
where $P_i(z)$ are $k$th order polynomial of $z$. The meromorphic
function is not holomorphic at poles as
\begin{equation}\label{2dgreen}
  \partial_{\bar{z}} \frac{1}{z} = 4\pi \delta^{2}(z).
\end{equation}
However, $w$ blows up at poles and so the self-dual equation
(\ref{Weq}) still holds, making the solutions (\ref{type1}) the
most general one.

For the commutative case, the solution (\ref{Weq}) is equivalent
to the smooth solution
\beq \Phi = (P_1(z),P_2(z),\cdots, P_{n+1}(z)) \frac{1}{1 + \sum_i
P_i^\dagger P_i}
\label{type2} \eeq
by a singular $U(1)$ gauge transformation $|P_{n+1}|/P_{n+1}(z)$.
In this case the vector
\beq
W = (P_1(z),P_2(z),\cdots,P_{n+1}(z))
\label{Wsol}
\eeq
has components which are $k$th order polynormials of $z$ only.
This solution has $2(n+1)k + 2n$ real parameters, among which $2n$
are the vacuum moduli parameter for $CP(n)$ space and the rest of
which $2(n+1)k$ parameters account for the size and scale
parameters of $k$ solitons. Note that the $W$ vector in
(\ref{Wsol}) is holomorphic everywhere.

When we goes to the noncommutative case, we should be more
careful. As $z^{-1} = (\bar{z}z)^{-1} \bar{z} = \bar{z} (\bar{z}z
+\theta)^{-1} $, we get
\beq
z z^{-1} = I, \;\; z^{-1} z = I-|0><0| .
\eeq
Since  $\partial_{\bar{z}} f(z,\bar{z}) = \theta^{-1}[z,f(z,\bar{z})]$,
\beq
\partial_{\bar{z}} z^{-1} = \theta^{-1} |0><0|,
\eeq
and the solution of type $1/z$ is not holomorphic on
noncommutative space. This has the analogue of (\ref{2dgreen}) on
noncommutative space.

In addition, we will see later that the solution $1/z$ will have
fractional topological charge. On the commutative case, two types
of solutions (\ref{type1}) and (\ref{type2}) are gauge equivalent,
but that is not true in general on the noncommutative case. While
$z^{-1}$ is nonholomorphic, the solutions given in
Eq.~(\ref{type2}) are polynomial so holomorphic, and so they are
solutions of the self-dual equation. This is the most general
solution even in the noncommutative space, modulo the local
scaling we considered before. Not only they satisfy the self-dual
equation (\ref{Weq}), these solutions also carry the integer
topological numbers.

Let us start with a simple  solution in the $CP(1)$ model,
\beq
\label{tn1}
W= ( az,1) ,
\eeq
and so
\beq
\partial_{\bar{z}} W^\dagger\,  P \, \partial_z W
= \frac{a^2}{1+a^2(\bar{z}z +\theta)} ,
\eeq
where we have used $z f(\bar{z}z) = f(\bar{z}z+\theta)z$
and $\bar{z}f(\bar{z}z) = f(\bar{z}z-\theta)\bar{z}$.
Thus the topological charge is
\beqn
Q_s &=& \frac{1}{2\pi} {\rm Tr}\left\{ \frac{1}{1+a^2\bar{z}z}
\left( \partial_{\bar{z}} W^\dagger\, P \,
\partial_z W   \right) \right\} \n\\
&=& \frac{1}{2\pi} {\rm Tr} \left\{
\frac{a^2}{(1+a^2\bar{z}z)\Bigl(1+a^2(\bar{z}z +\theta)\Bigr)}
\right\}.
\eeqn
With the dimensionless parameter $s = a^2\theta$, the trace becomes
\beqn
Q_s &=& \frac{1}{2\pi}(2\pi \theta) \sum_{n=0}^\infty
\frac{a^2}{ (1+ a^2\theta n)(1+ a^2\theta (n+1))} \n \\
&=&  s\sum \frac{1}{(1+sn)(1+s(n+1))} \n \\
&=& \sum_{n=0}^\infty \left( \frac{1}{1+sn} - \frac{1}{1+s(n+1)}
\right)= 1. \eeqn
The scale parameter $a$ of the soliton can be arbitrary but the topological
number does not change. Especially for the zero size soliton $a=\infty$,
topological charge density does not vanish only for the $|0>$ state.

If we have used the unacceptable singular solution
\beq W= (z^{-1}, 1) , \eeq
then its topological charge becomes
\beqn Q_s &=&\sum_{n=0}^\infty\{ 1/(1+s(n+1)) - 1/(1+s(n+2))\} \n\\
 &=& 1/(1+s) . \eeqn
As we argued before, this solution is not acceptable. Notice that
one can see that this solution has the topological charge less
than 1. This fractional topological number contrasts with the
commutative case. As on the noncommutative case $z^{-1}$ is as
singular as $z$ in the operator sense, one see that there can be a
configuration with a fractional topological charge. One thus
wonder whether one should include this configuration in the family
of classically acceptable configurations.  We do not know the
answer for this. This may be answerable by considering what kind
of soliton and anti-soliton pairs are created when some amount of
energy is put to the vacuum.

For a single anti-soliton solution in $CP(1)$ model,
\beq
\label{tn-1}
W = (a\bar{z},1),
\eeq
we get
\beq
\partial_z  W^\dagger \, P\,  \partial_{\bar{z}} W
= \frac{a^2}{1+a^2\bar{z}z}\;\; .
\eeq
Its topological charge is then
\beqn
Q_{\bar{s}} &=&- \frac{1}{2\pi}
{\rm Tr} \left\{ \frac{a^2}{(1+a^2(\bar{z}z +\theta))(1+a^2 \bar{z}z )}
\right\}\n \\
&=&-  s  \sum_{n=0}^\infty  \frac{1}{(1+sn)(1+s(n+1))} = - 1\;\; ,
\eeqn
with $s=a^2\theta$.

Thus the topological index works fine for these solitons
(\ref{type2}). This suggests that the topological charge can be
calculated for arbitrary (anti-)self-dual solutions. This is
indeed true as we will see now. For the general solution of
Eq.~(\ref{Wsol}), we can say
\beq
W = az^k u + {\cal O}( z^{k-1}),
\eeq
where $u$ is a $n+1$ dimensional unit vector.  To calculate its
topological charge, we first note that Eq.(\ref{Q(W)}) can be
rewritten as
\beq
Q_s = \frac{1}{2\pi} {\rm Tr} \left\{ \sqrt{W^\dagger W}
\partial_{\bar{z}}\left(
\frac{1}{W^\dagger W} W^\dagger \partial_z W \right)
\frac{1}{\sqrt{W^\dagger W}} \right\}.
\eeq
Now we can insert the complete set of states between operators to get
\beqn
Q_s &=& \theta \sum_{n,m,l} <n|\sqrt{W^\dagger W}|m>
<m| \partial_{\bar{z}} \left\{ \frac{1}{W^\dagger W} \partial_z(W^\dagger W)
\right\} |l>
<l|\frac{1}{\sqrt{W^\dagger W}} |n> \n\\
&=& \theta \sum_{n=0}^\infty
<n |\partial_{\bar{z}}\left(\frac{1}{W^\dagger W}
 W^\dagger \partial_z W \right)|n> \; .
\label{ncharge}
\eeqn
This is the analogue of the total derivative on  noncommutative plane.
Noting $\theta \partial_{\bar{z}} {\cal O}(z,\bar{z}) = [z, {\cal O}]$,
we can find the integration of the total derivative as \ct{gn}
\beqn
\frac{1}{2\pi} {\rm Tr} \partial_{\bar{z}} {\cal O} &=&
\sum_{n=0}^\infty  <n|[z,{\cal O}]|n>\n \\
&=& \sum_{n=0}^\infty \left\{ \sqrt{\theta (n+1)}<n+1|{\cal O}|n>
-\sqrt{\theta n}<n|{\cal O}|n-1> \right\} \n \\
&=&\lim_{N\rightarrow\infty} \sum_{n=0}^N \left\{  <n+1|{\cal O}z|n+1>
-<n|{\cal O}z|n> \right\} \n\\
&=& \lim_{N\rightarrow \infty} <N+1|{\cal O}z|N+1> \; ,
\label{bd-integral}
\eeqn
assuming that ${\cal O}z|0>=0$.  If ${\cal O}$ is singular and so
that ${\cal O}z|0>\neq 0$, there would be additional boundary
terms. (For example, the singular $w=u/(az^k)$ solution there is
an additional boundary term.) If the large $N$ limit vanishes, say
like $1/N$, then there is no boundary term and so the sum
vanishes. If the limit is of order one, the limit is finite. If it
diverges like a power of $N$, then the limit is not well defined.
In this case, the series should be treated more carefully.

Using the method in (\ref{bd-integral}), we get
\beq
Q_s =  \sum_{n=0}^\infty \left\{
<n+1| \frac{1}{W^\dagger W}  W^\dagger (\partial_z W) z |n+1> -
 <n|\frac{1}{W^\dagger W} W^\dagger (\partial_z W)z |n>\right\}.
\eeq
The expectation $<N+1|(W^\dagger W)^{-1}W^\dagger (\partial_z W)z|N+1>
$ is of order one.  More concretely, we see that
\beq
\frac{1}{W^\dagger W}W^\dagger (\partial_z W)z  =
\frac{1}{W^\dagger W}(W^\dagger (\partial_z W)z -kW^\dagger W)
+k.
\eeq
Defining that
\beq
\Omega \equiv \frac{1}{W^\dagger W} \left( W^\dagger (\partial_z W) z -
 k W^\dagger W \right) \; ,
\eeq
we see
\beq
\lim_{N\rightarrow \infty} <N+1|\Omega|N+1> = \lim_{N\rightarrow \infty}
 \frac{N^{k-1}}{N^k} =0 \; ,
\eeq
as  $ W^\dagger (\partial_z W) z -  k W^\dagger W
= {\cal O}((\bar{z}z)^{k-1}).$
Thus the charge becomes
\beqn
Q_s &=& \lim_{N\rightarrow \infty} \left\{ <N+1|\Omega|N+1>+ k \right\} \n\\
&=&  k.
\eeqn

For general anti-self-dual soliton solution,
\beq
W= a\bar{z}^k u + {\cal O}(\bar{z}^{k-1}),
\eeq
similar argument leads to the topological charge $-k$.

\section{Concluding Remarks}

In this Letter we have shown that the $CP(n)$ model can be also
well-defined on noncommutative two plane. There exist the
(anti-)self-dual solitons that saturate the BPS energy bounds,
which are regular and carry integer topological charge $k$ with
$2(n+1)k+2n$ real parameters. We found that (anti-)self-dual
solitons carry integer topological charge regardless their
orientation and size when the field configurations are regular. We
have also shown that the singular solutions, which are acceptable
and related to the regular solutions by gauge transformations on
the commutative plane, are not acceptable on noncommutative plane.
Not only they do not satisfy the self-dual equations, but also are
not related to the regular solution by the gauge transformation on
the noncommutative plane.  As we have seen, the topological number
does not change when the soliton shrinks to zero size. This should
remain true after going to the commutative variables by using the
Seiberg-Witten map \ct{sw}. While it is not clear how that is
achieved in our case, there is no natural way to evoke something
like freckled instantons and make two space to blow up at some
points, contrast to Braden and Nekrasov's work \ct{bn}.

The low energy dynamics of the solitons will be described on the
moduli space. To do this, one has to know the metric of the moduli
space \ct{ward}.  Our general solutions for $k$ solitons has
$2(n+1)k+ 2n$ real parameters in $CP(n)$ model. The vacuum moduli
space has $2n$ real parameters and their kinetic energy diverges
due to the volume factor. In addition, the total scale and
orientations with $2n$ real parameters have infinite inertia.  So
a single soliton with $k=1$ has only two parameters with finite
inertia, corresponding to the position of the soliton. For $k$
solitons, the moduli space ${\cal M}_{k,n}$ of finite inertia has
$2(n+1)k - 2n$ real dimension. The low energy dynamics of these
$k$ solitons can be described by the metric of the moduli space.
The solitons would not feel the noncommutativity of the underlying
space directly. However, the moduli space of solitons on
noncommutative space would be no longer singular when a single
soliton collapses to a point. It would be interesting to study in
detail the moduli space dynamics of solitons and compare that with
those on the commmutative plane.

The solitons in the gauge theories are more complicated than the
scalar theory like $CP(n)$.  Recently there has been many works on the
instantons on the noncommutative ${\bf R}^4$
\ct{ns,ky,tera,bn,furu,kly}.  The soliton properties of the
noncommutative $CP(n)$ model may play an important role in figuring
out the subtle issues in the noncommutative  instantons of four dimensional
Yang-Mills theory.  One of the key observation from our work is that
the topological number on the noncommutative space is somewhat tricky
quantity, which needs a careful treatment. Clearly one see a possible
solution to the recently discovered quandary \ct{kly} where the
instanton number of a single $U(2)$ instanton depends on the size of
the instanton.

{\bf Acknowledgment} $\;\;$ BHL is supported by the Ministry of
Education, BK21 Project No. D-0055 and by grant No. 1999-2-112-001-5
from the Interdisciplinary Research Program of the KOSEF and would
like to thank the hospitality in KIAS where the part of this work is
done.  KL is supported in part by KOSEF 1998 Interdisciplinary
Research Grant 98-07-02-07-01-5 and would like to thank the
hospitality in Niels Bohr Institute where the part of this work is
done.

\end{document}